# Quantum key distribution based on mid-infrared and telecom Band two-color entanglement source

Wu-Zhen Li[1,2,3,5], Chun Zhou[4,5], Yang Wang[4,5], Li Chen[1,2,3], Ren-Hui Chen[1,2,3], Zhao-Qi-Zhi Han[1,2,3], Ming-Yuan Gao[1,2,3], Xiao-Hua Wang[1,2,3], Di-Yuan Zheng[4], Meng-Yu Xie[1,2,3], Yin-Hai Li[1,2,3], Zhi-Yuan Zhou[1,2,3*], Wan-Su Bao[4*], and Bao-Sen Shi[1,2,3*]

[1] CAS Key Laboratory of Quantum Information, University of Science and Technology of China, Hefei, Anhui 230026, China

[2] CAS Center for Excellence in Quantum Information and Quantum Physics, University of Science and Technology of China, Hefei 230026, China

[3] Hefei National Laboratory, University of Science and Technology of China, Hefei 230088, China

[4] Henan Key Laboratory of Quantum Information and Cryptography, SSF IEU, Zhengzhou 450001, China

5 These authors contributed equally to this work

*zyzhouphy@ustc.edu.cn

*bws@qiclab.cn

* drshi@ustc.edu.cn

## ABSTRACT

Due to the high noise caused by solar background radiation, the existing satellite-based free-space quantum key distribution (QKD) experiments are mainly carried out at night, hindering the establishment of a practical all-day real-time global-scale quantum network. Given that the 3-5 μm mid-infrared (MIR) band has extremely low solar background radiation and strong scattering resistance, it is one of the ideal bands for free-space quantum communication. Here, firstly, we report on the preparation of a high-quality MIR (3370 nm) and telecom band (1555 nm) two-color polarization-entangled photon source, then we use this source to realize a principle QKD based on free-space and fiber hybrid channels in a laboratory. The theoretical analysis clearly shows that a long-distance QKD over 500 km of free-space and 96 km of fiber hybrid channels can be reached simultaneously. This work represents a significant step toward developing all-day global-scale quantum communication networks.

## INTRODUCTION

Entanglement is the foundation of many quantum information processing tasks, including quantum communication (*1, 2*), quantum computing (*3, 4*), quantum precision measurement (*5, 6*), and imaging (*7, 8*). Particularly in quantum key distribution (QKD), entanglement-based protocols (*9, 10*) provide certain advantages over prepare-and-measure protocols (*11, 12*), such as device-independent security (*13, 14*), quantum repeater compatibility (*15*) and anti-eavesdropping security (*16, 17*).

In recent years, entanglement-based QKD has been successfully implemented in metropolitan quantum networks (*18*), a satellite-relay intercontinental quantum network (*19*), and an integrated space-to-ground quantum network based on free-space and fiber hybrid channels (*20*), using the entangled photons in the near-infrared (NIR) band or in the telecom band, which proves the feasibility of entanglement-based QKD in building a global-scale quantum communication network. However, due to the high noise caused by solar background radiation (*21, 22*), existing satellite-based free-space QKD experiments can only be carried out at night, hindering the establishment of a practical all-day real-time global-scale quantum network. Recently, some works (*23-27*) have demonstrated the workability of QKD in free space under daylight; for example, Liao *et al.* demonstrated free-space QKD over 53 km on sunny days, using frequency upconversion (FUC) technology to detect 1550 nm photons (*28*). Li *et al.* demonstrated a hybrid measurement-device-independent QKD network consisting of free-space and fiber channels in full daytime for the first time (*29*). However, in these works, strong filtering with spectral, temporal, and spatial filters is needed to eliminate noise. Besides, strong filtering typically introduces additional losses to the signal photons and also requires high-precision frequency alignment with the source.

The above works have primarily been done in the NIR band, there is neither any report on entanglement-based QKD in the mid-infrared (MIR) band till now, nor even QKD works by using prepare-and-measure protocol. Compared to the NIR band, the photons in the MIR spectral region of 3-5 μm, known as an atmospheric communication window, suffer lower solar background radiation and have stronger atmospheric penetration ability. For example, the solar background radiation and Rayleigh scattering loss at 3.37 μm are 16 times and 22 times lower than those at 1550 nm, respectively (*30, 31*), making it particularly promising for free-space quantum communications during daytime. However, some practical factors limit the development of this work. On the one hand, single-photon detectors in the MIR band are currently relatively immature. The commonly used semiconductor detectors and superconducting nanowire single-photon detectors face challenges in improving detection efficiency and signal-to-noise ratio. Moreover, these detectors require deep cooling to avoid the impact of thermal noise, which makes the entire detection system more complex and expensive (*32, 33*). A practical solution is to convert MIR photons into visible or NIR bands through the FUC process, which can then be detected using mature silicon-based detectors (*34-38*). Numerous studies have shown that the FUC process can preserve the quantum properties of the photons and significantly reduce the impact of environmental thermal noise (*39-41*). On the other hand, although many technological advances have been made in MIR band communication fibers (*42-44*), they have not yet been widely used.

In this work, we report for the first time a two-color polarization entanglement source at 1555 nm and 3370 nm via the type-0 spontaneous parametric down-conversion (SPDC) process by using two orthogonally placed MgO-doped periodically poled lithium niobate (MgO: PPLN) crystals pumped by a 1064 nm pulse laser. After that, the MIR idler photons are upconverted to 808.7 nm through the FUC process for efficient detection. The average visibility of two-photon interference is about 95.8% ± 0.4%. The measured Clauser-Horne-Shimony-Holt (CHSH) inequality $S$ parameter is 2.7144 ± 0.0068, which violates the inequality with 105 standard deviations. The fidelity of the generated entangled state is 0.961 ± 0.0027. Furthermore, we perform a principle entanglement-based QKD experiment through a hybrid channel in the laboratory, where the signal photons and idler photons were transmitted in a 5 m-long fiber channel and a 3.5 m-long free-space channel, respectively. The obtained secure key rate is 65 bits·s$^{-1}$, and the average quantum bit error rate (QBER) is 4.4%. In addition, the long-distance transmission capability of the QKD system over 500 km of free-space and 96 km of fiber hybrid channels is theoretically verified. This work paves the way for all-day communications in a global-scale quantum network based on free-space and fiber hybrid channels.

**RESULTS**

**Experimental setup**

As shown in Fig. 1, the experiment setup mainly consists of three parts: an entangled photon pair generation module, an FUC module, and a coincidence detection module. The two-color polarization-entangled photon pairs generated via the type-0 SPDC process in two crossed MgO: PPLN crystals are separated by a dichromatic mirror. The reflected 1555 nm signal photons are directly detected by an indium gallium arsenide avalanche photodiode (InGaAs-APD), while the transmitted 3370 nm idler photons are first upconverted to 808.7 nm by the FUC module and then detected by a silicon avalanche photodiode (Si-APD). The optical delay line (ODL) is used to precisely control the temporal overlap between the pump pulse and MIR photons in the FUC module. Full details of the experimental setup are summarized in the Materials and Methods section.

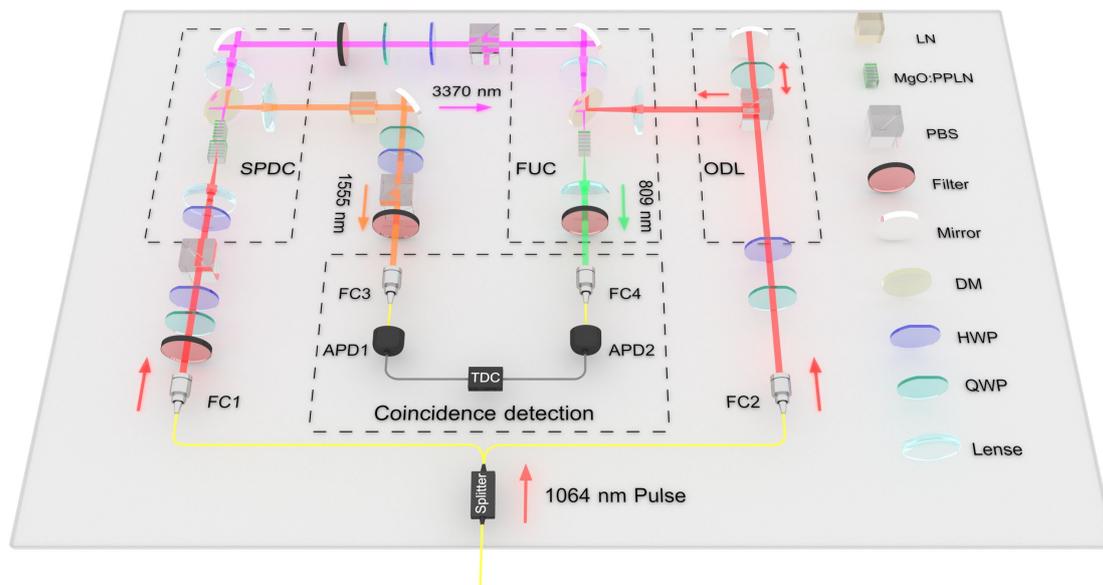

**Fig. 1. Experimental setup for preparation of two-color polarization-entangled photon pairs and characterization of entanglement.** FC, fiber collimator; APD, avalanche photodiode; TDC, time-to-digital converter; MgO: PPLN, MgO-doped periodically poled lithium niobate; LN, lithium niobate; PBS, polarization beam splitter; DM, dichromatic mirror; HWP, half-wave plate; QWP, quarter-wave plate;

**The performance of the FUC module**

To gain insight into the detection efficiency of MIR photons, we evaluate the working performance of the FUC module by testing the quantum conversion efficiency (QCE). Here, the 3370 nm MIR classic light source produced through a difference frequency generation process is used as a substitute for MIR photons. To ensure the non-depleted pump approximation is fulfilled, the power of the MIR source is set to 0.1 mW. The optimal QCE under different pump powers, as shown in Fig. 2A, is obtained by precisely adjusting the spatial and temporal overlap between the MIR pulse and the 1064 nm pump pulse in the nonlinear crystal. We achieve a QCE of nearly 50% or a power efficiency of ~200% when the pump power is increased to 1.5 W, which is still some way from the saturation state. In addition, the optimal phase-matching temperatures of the crystals in the SPDC and the FUC processes can be determined in this way. The estimated acceptance bandwidth of the FUC module is about 4.8 nm, which can be extended by using a shorter nonlinear crystal at the expense of conversion efficiency. In addition, the detection of wide-spectrum MIR photons can be achieved by scanning the poling period and phase-matching temperature of the nonlinear crystal (*41*). The more experimental details are presented in the Supplementary Material, Sec. S1.

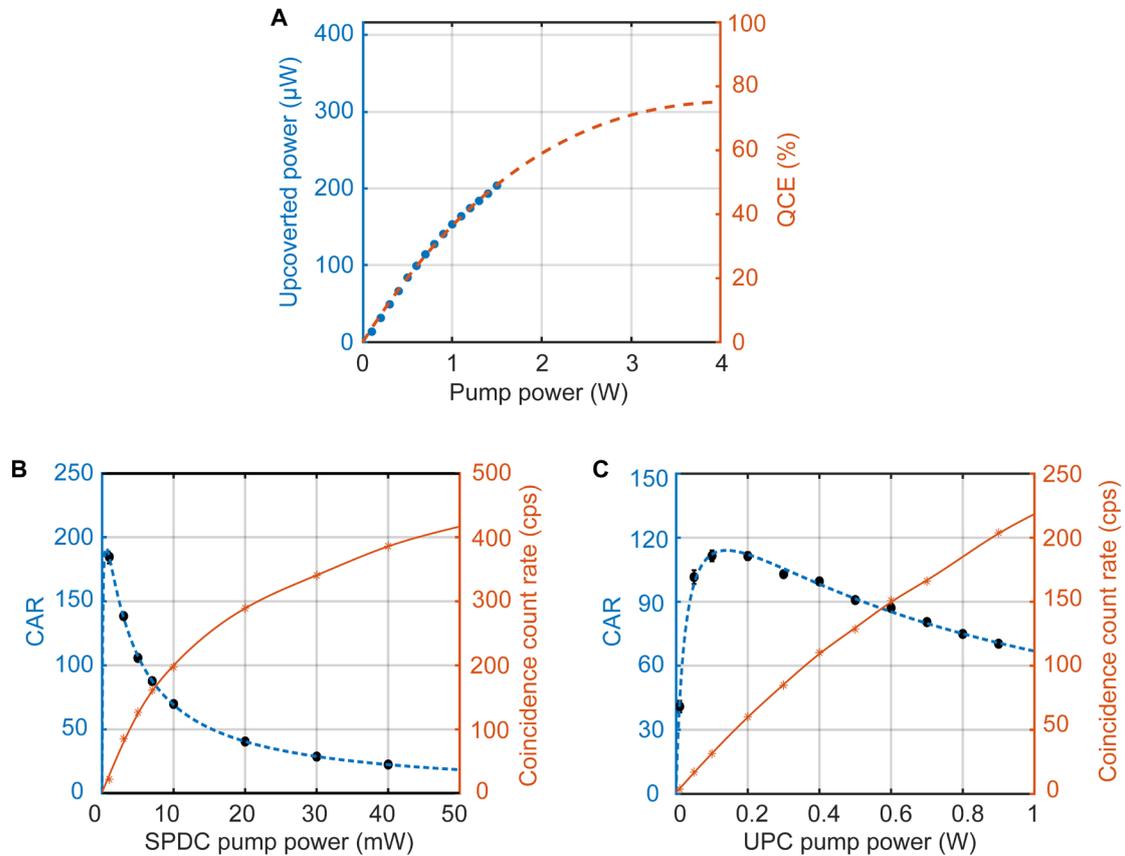

**Fig. 2. Performance of the FUC module and coincidence measurement of two-color photon pairs.** (**A**) Measured 808.7 nm upconverted power/QCE as a function of the 1064 nm pump power. The dashed line is given by the theoretical model in (*35*). (**B**) Variation of CAR and coincidence count rate with SPDC pump power. The pump power in the FUC module is set to 0.4 W. (**C**) Variation of CAR and coincidence count rate with FUC pump power. The pump power in the SPDC process is 4 mW. The accumulation time for coincidence measurement is 60 s, and the coincidence window is 3.2 ns.

## Coincidence measurement of the two-color photon pairs

Coincidence-to-accident ratio (CAR) is a crucial metric for assessing the quality of an entanglement source. A high CAR value indicates that the entangled state has high purity and low noise, thus ensuring the reliability and security of quantum communication. To achieve the optimal balance between state purity and count rate, we project the entangled state into the $|VV\rangle$ and then perform measurements of the CAR at various pump powers. As shown in Fig. 2B, the CAR value first increases and then decreases with the increase of SPDC pump power. This is because excessive pump power increases the probability of generating multiple photon pairs per pulse *(45)*. We also measured the CAR at different FUC pump powers, as shown in Fig. 2C. At low pump powers, the noise mainly originates from the detector itself and the background. As the pump power increases, noise caused by unintended nonlinear optical processes begins to dominate *(46, 47)*. Although the maximum CAR we measured is only 185 with an SPDC pump power of 1 mW and a FUC pump power of 0.4 W, we believe that the maximum CAR value will exceed 200 by optimizing the FUC pump power.

## Characterization of the polarization entanglement

As shown in Fig. 1, the diagonally polarized pump pulse passes through two crossed MgO: PPLN crystals and the maximally entangled state generated by two equally probable type-0 SPDC processes can be expressed as

$$|\Phi^\varphi\rangle = \frac{1}{\sqrt{2}}\left(|H_s H_i\rangle + e^{i\varphi}|V_s V_i\rangle\right), \qquad (1)$$

The relative phase $\varphi$ between $|H_s H_i\rangle$ and $|V_s V_i\rangle$ can be adjusted by tilting the temporal compensation crystal LN. Here, $\varphi$ is set to $\pi$, the polarization-encoded Bell state we obtain can be written as

$$|\Phi^-\rangle = \frac{1}{\sqrt{2}}\left(|H_s H_i\rangle - |V_s V_i\rangle\right), \qquad (2)$$

The SPDC pump power is set to 10 mW, and the FUC pump power is fixed at 0.5 W. To measure the polarization correlations between signal and idler photons, the HWP in the idler path is set to 45°, 0°, 22.5° and -22.5° to project the signal photon into the four polarization states represented by $|H\rangle$, $|V\rangle$, $|D\rangle$, and $|A\rangle$, respectively, and the polarization curves, as shown in Fig. 3A, are obtained by rotating the HWP in the signal path. Without correcting for accidental coincidence counts, the visibilities in the H, V, D, and A bases calculated by sinusoidal function fitting are 97.3% ± 0.3%, 97.4% ± 0.3%, 94.8% ± 0.4%, and 93.5% ± 0.5%, respectively. As described in the Supplementary Material, Sec. S2, the lower visibility in the A and D bases is attributed to the spatial walk-off of the orthogonally polarized signal photons caused by the non-ideal temporal compensation crystal and the slight difference between the two MgO: PPLN crystals used for SPDC. Furthermore, we also perform the CHSH-Bell inequality test (*48*) and obtain S=2.7144 ± 0.0068 in 60 s, which violates the inequality by 105 standard deviations and agrees with the expected value $S = 2\sqrt{2}V$, where $V$ is the average visibility.

To assess the degree of entanglement more completely, we perform quantum state tomography on the entangled photon pairs (*49*). The density matrix $\rho_{\text{exp}}$ of the generated entangled state is reconstructed using the maximum likelihood estimation method, as shown in Fig. 3B, and the fidelity relative to the pure Bell state $\left|\Phi^-\right\rangle$, defined as $F = \left\langle\Phi^-\right|\rho_{\text{exp}}\left|\Phi^-\right\rangle$, is found to be 0.961 ± 0.0027 for the generated state. In order to obtain high fidelity, we precisely calibrate the optical axis angle of wave plates before measurement and use a wavelength-insensitive true zero-order wave plate in the wide-spectrum MIR optical path. Besides, the pump pulse is output from a single-mode fiber, which can ensure a higher beam quality and improve the fiber collection efficiency of photon pairs. These results, obtained without subtracting the accidental coincidence counts, indicate that our entanglement source produces very high-quality polarization-entangled photon pairs, which is very important for QKD systems that do not allow background subtraction.

In addition, we also estimate the spectral brightness of the entanglement source. The estimated effective detection efficiencies (including collection and detection efficiencies) of signal and idler photons are 0.04 and 0.0038, respectively, of which the detection efficiency of idler photons is 0.08 (including the QCE of the FUC module of 0.2, the coupling efficiency of 0.67, and the detection efficiency of the Si-APD of 0.6). Therefore, the emission spectral brightness is 0.17 (s·mW·MHz)$^{-1}$. (see Supplementary Material, Sec. S3 for more details).

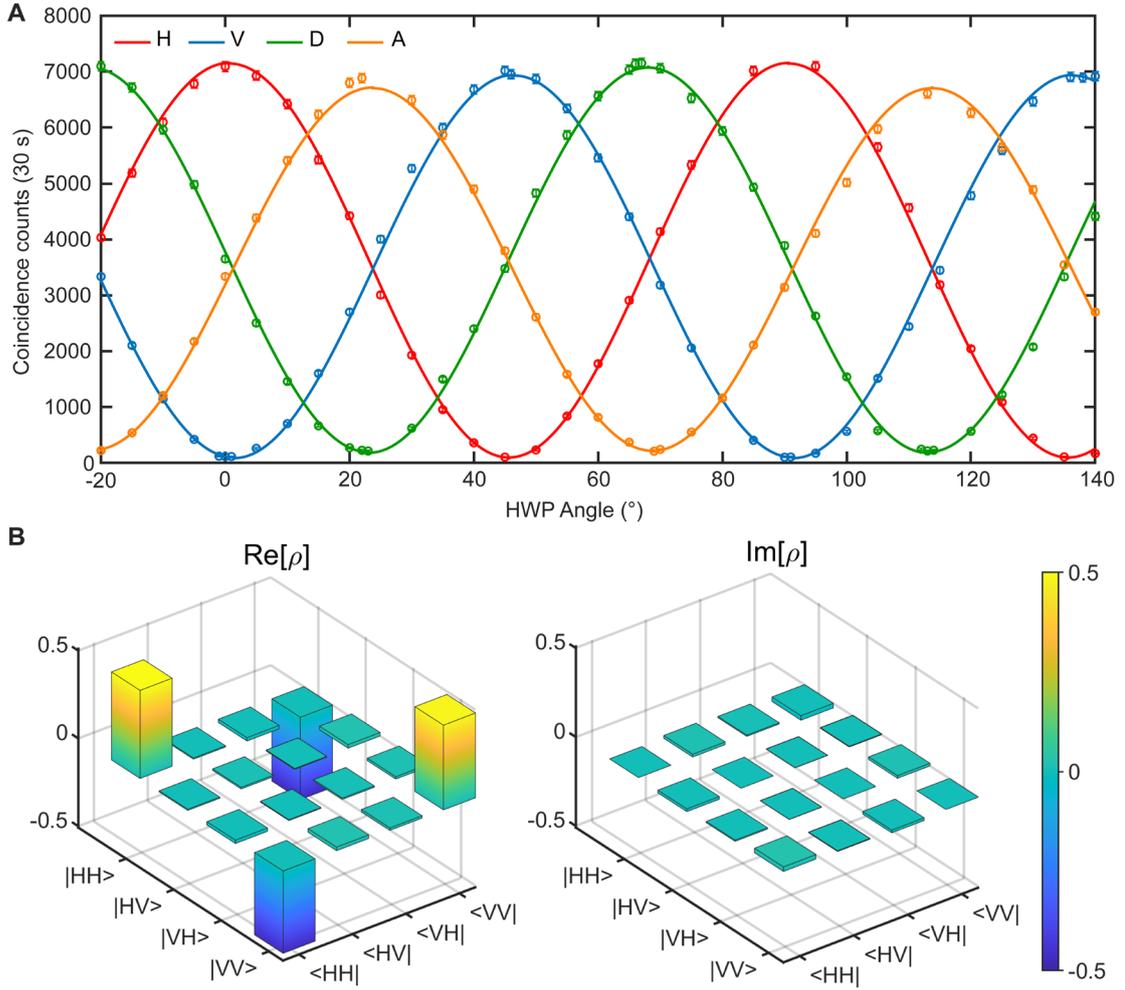

**Fig. 3. Polarization entanglement characterization of the generated entangled state. (A)** Polarization interference in four bases; **(B)** the real (left) and imaginary (right) parts of the density matrix reconstructed by the maximum likelihood estimation method. The above results are obtained with a coincidence accumulation time of 30 s and a coincidence window of 3.2 ns.

### Entanglement-based QKD for hybrid channels

Given that the source operates in the MIR and telecom bands, it is inherently suitable for free-space communications during daylight and fully compatible with telecom-band fiber networks. We perform a QKD experiment based on the BBM92 protocol (*10*) in the laboratory to more clearly demonstrate the potential of this source in quantum communication. As shown in Fig. 4, the two-color polarization-entangled photon pairs prepared from the entanglement source are split into two paths, where the 1555 nm signal photons are transmitted to Alice via a 5 m single mode telecom fiber (SMF28-e, ITU G652D; mode field diameter MFD of 9.2 μm), and the 3370 nm idler photons are transmitted to Bob via a 3.5 m free-space. The pump powers of the entanglement source and FUC module are 10 mW and 700 mW, respectively, and the time window is set to 3.2 ns. After processing the photon event data recorded by Alice and Bob, we obtain a raw key rate of 306.3 bits·s$^{-1}$ and a shifted key rate of 153 bits·s$^{-1}$, which can be further improved by increasing the SPDC pump power or optimizing the coupling efficiency.

The average QBER is 4.4%, with a QBER of 3.6% on the H/V basis and 5.3% on the M/P basis, all below the 11% security limit (*50*). The secure secret key rate estimated by the formula given in (*17*) is 65 bits·s$^{-1}$.

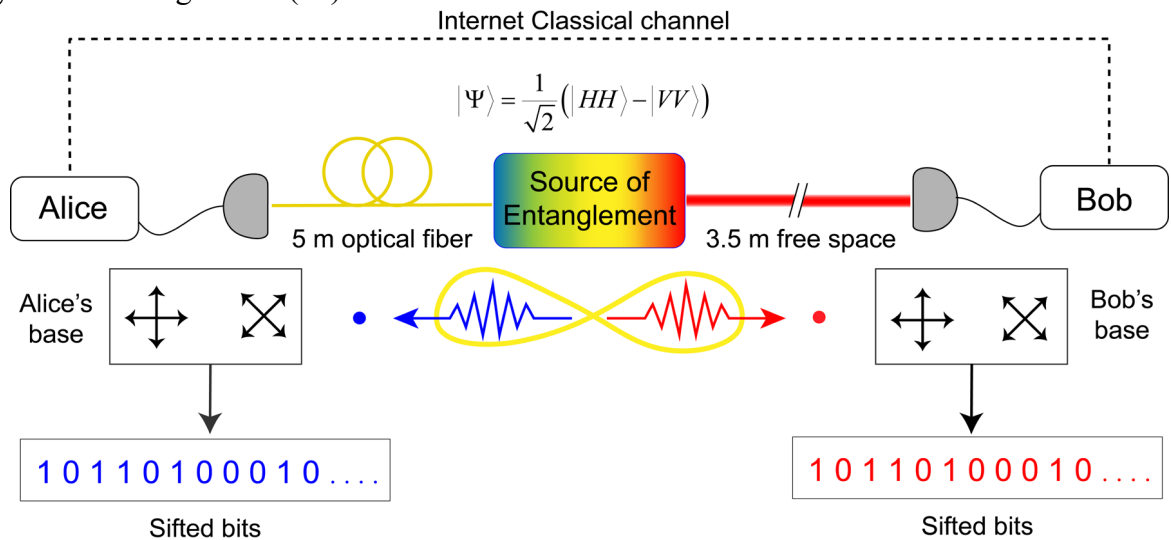

**Fig. 4. Schematic diagram of entanglement-based QKD experiment for free-space and fiber hybrid channels.** Alice and Bob receive entangled photon pairs from the entanglement source via fiber and free-space channels, respectively, and perform measurements on them. Then, Alice and Bob exchange their chosen measurement bases through a classical channel and sift out the keys with matching measurement bases from the raw key. Finally, they select a portion of the sifted key to estimate the QBER.

In addition, based on the above experimental parameters of the QKD system, we model the secret key rate (SKR) using the theoretical results in (*17, 51, 52*) as a function of the lengths of the free-space and fiber channels by using solar spectral irradiance with a 5nm bandwidth around the free-space carrier wavelength. As shown in Fig. 5, the QKD system can maintain a positive SKR after transmission through 500 km of free space and 96 km of fiber simultaneously, demonstrating significant long-distance communication capability during daylight. Moreover, the communication distance of the QKD system can be further improved by improving the yield of entangled photon pairs and the detection efficiency of MIR photons. (Supplementary Material, Sec. S4 describes more details)

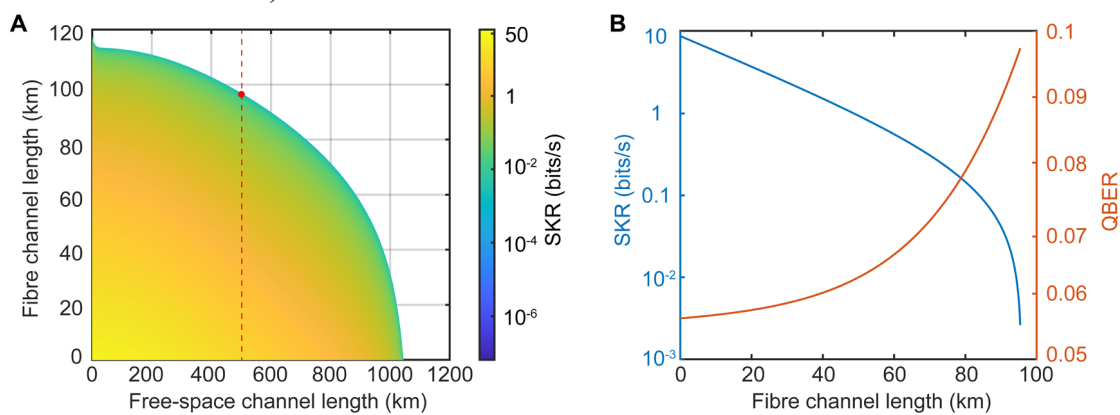

**Fig. 5. Transmission characteristics of the QKD system in free-space and fiber hybrid channels.**

(**A**) The secret key rate (SKR) of the QKD system varies with the length of the free space and fiber channels. (**B**). At a 500 km free-space channel length (indicated by the dashed line in (**A**)), the SKR and quantum bit error rate (QBER) as a function of fiber channel length are analyzed.

## DISCUSSION

We have developed a two-color polarization entanglement source at 3370 nm and 1555 nm based on two crossed MgO: PPLN crystals. The high entanglement quality of the source has been verified by multiple characterization methods. Furthermore, through the QKD demonstration experiment, we have shown the potential and practicality of this source for long-distance quantum communication over hybrid channels. Given its advantages in free-space communication and compatibility with telecom-band fiber networks, this entanglement source stands out as a promising candidate for an all-day, global-scale hybrid quantum network. In the future, we will explore the application of this entanglement source in long-distance quantum communication during daylight. Furthermore, this work proposes an efficient MIR single-photon detection scheme based on the FUC process with the advantages of operating at room temperature and real-time detection. Moreover, the response band can be easily tuned by changing the length, polarization period, and temperature of the nonlinear crystal. Our results bring more opportunities for advanced quantum technology applications in the MIR spectral band.

## MATERIALS AND METHODS
### Preparation of entangled photon pairs

As shown in Fig. 1, the 1064 nm pulsed light emitted from a high-power ytterbium-doped fiber laser with a repetition frequency of 100 MHz, a pulse duration of 100 ps, and average power of up to 2 W is split into two paths by a homemade beam splitter; one path is used as the pump for the SPDC process to prepare entangled photon pairs, and the other path is used as the pump for the UCP module. A band pass filter (BPF) with a bandwidth of 10 nm centered at 1064 nm and a polarization beam splitter (PBS) are used to "clean up" the spectrum and polarization of the SPDC pump pulse. The two nonlinear crystals used for the SPDC process are nearly identical type 0 phase-matched MgO: PPLN crystals, each with a length of 4.5 mm, an aperture of 0.5 mm by 0.5 mm, and a poling period of 30.5 um. The two crystals are mounted orthogonally in a double oven consisting of two individual Peltier elements whose temperatures are precisely controlled at 75 °C using two homemade semiconductor temperature controllers (temperature stability of ±2 mK). For equal crystal excitation, the SPDC pump polarization is rotated 45° with respect to the crystal axes by an HWP, and the beam is focused at the boundary of two crystals by a 150-mm lens with a beam waist of 42 μm at the focus. The generated two-color polarization-entangled photon pairs are separated by a dichroic mirror (DM). The reflected 1555 nm signal photons are collimated by a lens with a focal length of 100 mm and then collected by a single-mode fiber. The transmitted 3370 nm idler photons are collimated by a $CaF_2$ lens with a focal length of

100 mm and then detected by the FUC module. A 5 mm long lithium niobate (LN) crystal (cut angle $\theta = 45.4°$) in the signal path is used as a temporal compensation crystal to eliminate the wavelength-dependent phase. In order to erase as much distinguishable information as possible, the temperature of the LN crystal is controlled at 32.5 °C. A filter set consists of two DMs and a band-pass filter with a bandwidth of 10 nm centered at 1555 nm, which is used to filter out residual pump noise in the signal path. The residual pump in the idle path is filtered out by a band-pass filter with a central wavelength of 3um and a bandwidth of 250 nm. A polarization analyzer consisting of an HWP, a QWP, and a PBS in each path is used for entanglement characterization measurements.

**Upconversion detection of MIR photons**
In the FUC module, a type 0 MgO: PPLN crystal is chosen that has a poling period of 22.4 μm, a length of 40 mm, and an aperture of 0.5 mm by 0.5 mm. The temperature of the crystal is stabilized at 78 °C. To achieve optimal QCE, the pump pulse used for the FUC module needs to overlap with the MIR photons in time before entering the crystal. Therefore, we constructed an optical delay line (ODL) consisting of a PBS, a QWP, and a mirror to precisely adjust the delay of the pump pulse by controlling the position of the mirrors. Two lenses with the same focal length of 150 mm are used to focus the MIR photons and 1064 nm pump pulses, respectively. Then, the converted 808.7 nm photons are collimated by a lens with a focal length of 150mm and then pass through a filter set consisting of a DM, a 750-nm long-pass filter, an 850-nm short-pass filter, and a band-pass filter with a bandwidth of 40 nm centered at 800 nm to filter out the residual pump. Finally, the converted 808.7 nm photons are coupled into a single-mode optical fiber with a coupling efficiency of about 67%.

**Coincidence detection**
The detector for 1555 nm signal photons is an InGaAs-APD (IDQ, ID220-FR-SMF) with a detection efficiency of 20% and a dead time of 5 μs. For the converted 808.7 nm photons, we use a Si-APD (Excelitas, SPCM-AQRH-11-FC) with a detection efficiency of approximately 60% and a dead time of 40 ns. Ultimately, the coincident measurement events are recorded via a fast time-to-digital converter (PicoQuant, TimeHarp 260) with the coincidence window set to 3.2 ns.

**Error estimation**
Since the Bell parameter $S$ is a function of 16 coincidence counts $C_j$, where the subscript $j = 1,…,16$ corresponds to all combinations of measurement angles, the standard deviation of the parameter $S$ can be expressed as

$$\sigma_S = \left( \sum_{j=1}^{16} \left( \sigma_{C_j} \frac{\partial S}{\partial C_j} \right)^2 \right)^{\frac{1}{2}}, \tag{3}$$

where $\frac{\partial S}{\partial C_j}$ is the partial derivative of $S$ with respect $C_j$, and $\sigma_{C_j} = \sqrt{C_j}$ is the uncertainty of $C_j$ given by the Poisson counting statistical. Similarly, we estimate the error bars by assuming a Poisson distribution of the data.

**Supplementary Materials**
This PDF file includes:
    Supplementary Text
    Figs. S1-S4
    Tables S1

**Acknowledgments**
**Funding:** we would like to acknowledge the support from the National Key Research and Development Program of China (2022YFB3607700, 2022YFB3903102), National Natural Science Foundation of China (NSFC) (11934013, 92065101, 62005068), and Innovation Program for Quantum Science and Technology (2021ZD0301100), and the Space Debris Research Project of China (No. KJSP2020020202), and USTC Research Funds of the Double First-Class Initiative. This work was supported by the Opening Funding of National Key Laboratory of Electromagnetic Space Security. **Author contributions:** Z.-Y.Z., W.-S.B., and B.-S.S. designed the experiment. W.-Z.L., C.Z., and Y.W. carried out the experiment. M.-Y.X and Y.-H.L. provided software support. L.C., R.-H.C., Z.-Q.-Z.H., X.-H.W., and M.-Y.G. helped collect the data. Z.-Y.Z. and W.-Z.L. analyzed the data and wrote the paper with input from all other authors. The project was supervised by Z.-Y.Z., W.-S.B., and B.-S.S. All authors discussed the experimental procedures and results. **Competing interests:** The authors declare no competing financial interests. **Data and materials availability:** All data needed to evaluate the conclusions in the paper are present in the paper and the Supplementary Materials.

# Supplementary Materials

# Quantum key distribution based on mid-infrared and telecom band two-color entanglement source

Wu-Zhen Li[1, 2, 3,5], Chun Zhou[4,5], Yang Wang[4,5], Li Chen[1, 2, 3], Ren-Hui Chen[1, 2, 3], Zhao-Qi-Zhi Han[1, 2,3], Ming-Yuan Gao[1, 2, 3], Xiao-Hua Wang[1, 2, 3], Di-Yuan Zheng[4], Meng-Yu Xie[1, 2, 3], Yin-Hai Li[1, 2, 3], Zhi-Yuan Zhou[1, 2, 3*], Wan-Su Bao*, and Bao-Sen Shi[1, 2, 3*]

[1] *CAS Key Laboratory of Quantum Information, University of Science and Technology of China, Hefei, Anhui 230026, China*
[2] *CAS Center for Excellence in Quantum Information and Quantum Physics, University of Science and Technology of China, Hefei 230026, China*
[3] *Hefei National Laboratory, University of Science and Technology of China, Hefei 230088, China*
[4] *Henan Key Laboratory of Quantum Information and Cryptography, SSF IEU, Zhengzhou 450001, China*
5 These authors contributed equally to this work

*zyzhouphy@ustc.edu.cn
*bws@qiclab.cn
* drshi@ustc.edu.cn

**S1. Characterization of the UPC module**

As mentioned in the main article, the mid-infrared (MIR) classical light used in the UCP module characterization experiment is obtained via the difference frequency generation process (DFG). The pump beam used for the DFG is also provided by the 1064 nm high-power pulsed laser mentioned in the main article. The signal beam at 1555 nm for the DFG comes from a tunable narrow-linewidth continuous laser and is amplified by a fiber amplifier. The spectra of the pump and signal beams are shown in Fig. S1(a) and (b). The nonlinear crystal used for the DFG is a type 0 MgO: PPLN crystal with a poling period of 30.5 μm, a length of 10 mm, and an aperture of 1 mm × 1 mm. The generated MIR laser power exceeds 1mW, and the evaluated bandwidth is 0.522nm. Then, the MIR beam is injected into the UPC module and interacts with the time-synchronized 1064 nm pump pulse in the crystal. The generated up-converted pulse is coupled into a single-mode fiber after filtering, with a filtering efficiency of 92% and a coupling efficiency of 67%. Due to the limitation of phase-matching conditions, the UCP module can only up-convert MIR photons within a limited spectral bandwidth. For a 40mm long PPLN crystal with a poling period of 22.5 μm used in the UCP module, its acceptance bandwidth is 4.8 nm. When the incident MIR photon spectrum is relatively wide, a short crystal (or chirped polarization crystal) with a larger acceptance bandwidth can be used, but this may lead to a decrease in quantum conversion efficiency (QCE); another method is to scan the temperature of the crystal to achieve UPC of the entire spectrum. As shown in Fig. S1(c), theoretical simulation shows the change of the acceptance bandwidth of the crystal used in the UPC module with temperature. It is worth emphasizing that the UPC process can effectively filter out the environmental thermal noise, thereby achieving a high signal-to-noise ratio at room temperature (*41*).

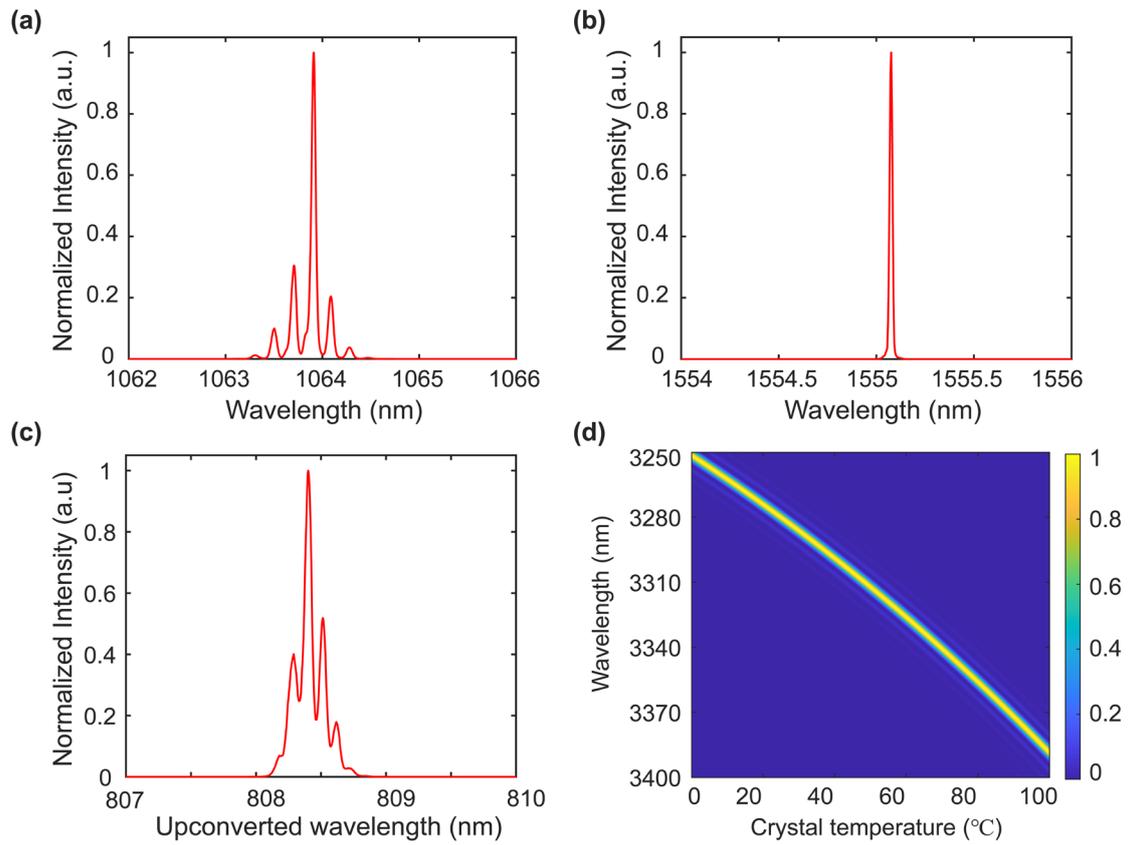

Fig. S1 (a)-(b) Output spectra of 1064nm pulsed laser and 1555nm continuous laser used in the DFG process. (c) The measured spectrum of the up-converted pulse. (d) The simulated curve of the acceptance bandwidth of the UPC module as a function of the crystal temperature.

## S2. Temporal compensation

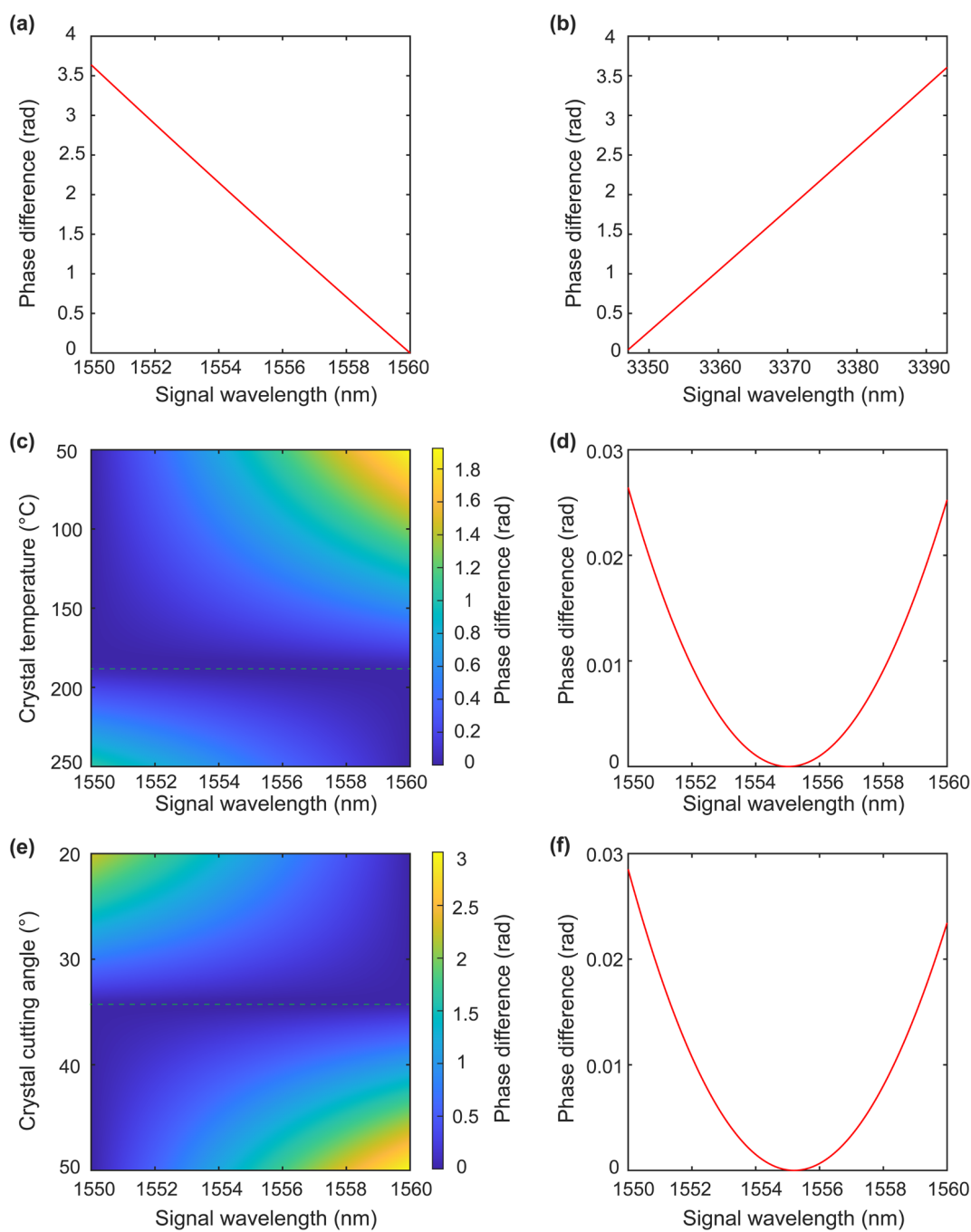

Fig. S2 Phase difference as a function of the wavelengths of the (a) signal and (b) idler photons before inserting the compensation crystal. Then, insert an LN crystal with a cutting angle of 45.4° and a length of 5 mm into the signal path. (c) The phase difference in the signal spectrum obtained by simulation varies with the crystal temperature. (d) A flat dependence can be obtained when the compensation crystal temperature is 187.5 °C (green dashed line in (c)). (e)-(f) The phase difference varies with the crystal cutting angle, where the green dashed line represents the optimal crystal cutting angle (34.2°) for temporal compensation, and the crystal temperature is set to 32.5 °C.

Since the entanglement source uses two crossed crystals to successively prepare non-degenerate photon pairs, the $|V_s V_i\rangle$ pairs generated in the first crystal acquire an additional phase shift relative to the $|H_s H_i\rangle$ pairs generated in the second crystal; in addition, the photon pairs generated by the SPDC process will pick up a wavelength-dependent phase after passing through the dispersive material, which leads to spectral distinguishability. Here we assume that the photon pairs are generated at the incident surface of the crystals, and the phase difference $\Delta\varphi$ between $|V_s V_i\rangle$ and $|H_s H_i\rangle$ pairs after passing through the crystal can be expressed as

$$\Delta\varphi(\lambda_s, \lambda_i) = \varphi_p^V + \varphi_s^V + \varphi_i^V - (\varphi_p^H + \varphi_s^H + \varphi_i^H) = 2\pi L\left(\frac{n_o(\lambda_s)}{\lambda_s} + \frac{n_o(\lambda_i)}{\lambda_i} - \frac{n_e(\lambda_p)}{\lambda_p}\right)$$

(1.1)

As shown in Fig. S2(a)-(b), in the absence of compensation, the phase difference $\Delta\varphi$ is severely dependent on the wavelengths of the signal and idler photons. In order to eliminate spectral distinguishability, a compensation crystal needs to be added to flatten this phase map. Here, we insert a 5mm long LN crystal (cutting angle $\theta = 45.4°$) in the signal path to compensate for the spectral phase, and the phase difference $\Delta\varphi_c$ after compensation is

$$\Delta\varphi_c = \Delta\varphi + \varphi_s^V - \varphi_s^H = \Delta\varphi + 2\pi L_{LN}\left(\frac{n_e(\theta, T)}{\lambda_s} - \frac{n_o(T)}{\lambda_s}\right) \qquad (1.2)$$

Therefore, the optimal time compensation can be achieved by adjusting the length $L_{LN}$, temperature $T$, and cutting angle $\theta$ of the compensation crystal [see (*53,54*) for more information]. As shown in Fig. S2(c)-(d), under the existing crystal conditions, the optimal crystal temperature simulated is 187.5°C, exceeding the maximum temperature the controller can reach. However, we can slightly tilt the compensation crystal (equivalent to changing the cutting angle of the crystal), as shown in Fig. S2(e)-(f), to achieve the optimal compensation.

Due to manufacturing constraints, the lengths of the crystals used for SPDC are slightly different, and the poling period in the crystal is also slightly non-uniform. In addition, the cutting angle of the time compensation crystal used in the experiment is not 90°, so the orthogonal polarization signal photons will be slightly separated in space due to the walk-off effect after passing through the LN crystal. These factors cause lower interference visibility on the A/D basis.

## S3. Brightness of the entanglement source

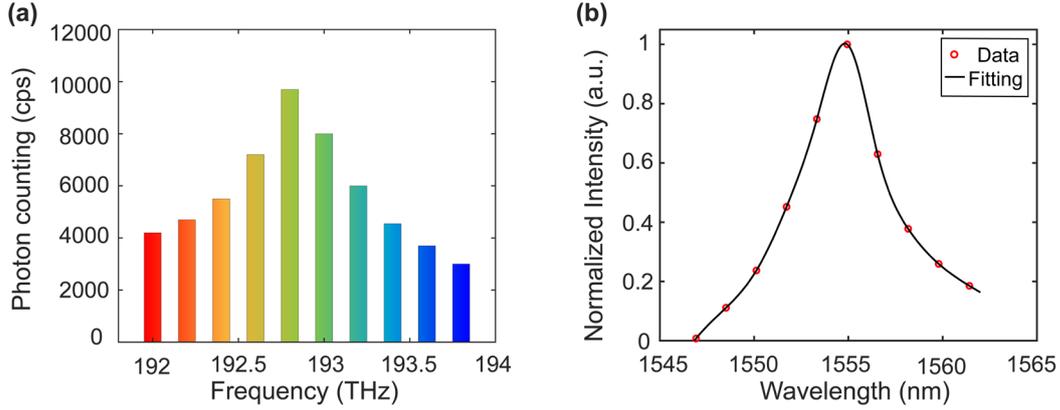

**Fig. S3 Spectrum of signal photons measured using a 100 GHz bandwidth DWDM.** (a) Each bar represents the photon count rate of a channel, and the width of the bar corresponds to the bandwidth of that channel. From left to right, the bars correspond to channels C20, C22, C24, C26, C28, C30, C32, C34, C36, and C38, respectively. (b) Spectrum of signal photons after subtracting accidental coincidence counts.

Before calculating the brightness of the entanglement source, it is necessary to evaluate the spectral bandwidth of the entangled photon pairs. We inject the collected signal photons into a Dense Wavelength Division Multiplexer (DWDM) with a bandwidth of 100 GHz and then measure the photon counts of each channel separately. As shown in Fig. S3(a), the highest photon count rate is observed in channel C28, corresponding to a wavelength of 1554.94 nm. The signal photon spectrum obtained after subtracting the accidental coincidence counts is shown in Fig. S3(b). The bandwidth of the signal photons is about 5.14 nm. The theoretically calculated spectral bandwidths of signal photons and idler photons are 15.4 nm and 72 nm, respectively. Optical filtering elements in the signal path are the main reason for the reduction of signal photon bandwidth. Similarly, the UPC module also filters the spectrum of the idler photons.

The collection efficiency of signal photons (including transmission, filtering, and coupling efficiency) $\alpha_1$ is 0.23, and the detection efficiency $\eta_1$ is 0.2. The collection efficiency of idler photons (including transmission and filtering efficiency) $\alpha_2$ is 0.0476, and the detection efficiency (including QCE of the UCP module of 0.2, coupling efficiency of 0.67, and detection efficiency of the silicon detector of 0.6) $\eta_2$ is 0.08. The spectral brightness of the entanglement source inferred from the

$$SB_{\text{Inferred}} = \frac{N_{CC}}{\alpha_s \alpha_i \eta_s \eta_i P \Delta v}$$ is about 0.17 (s·mW·MHz)$^{-1}$, where $Ncc$ = 496 cps is the

coincidence count rate, $\Delta v = 1.91 \times 10^6$ MHZ is the theoretical bandwidths of the signal and idler photons, and $P$ = 10 mW is the pump power. The detected spectral brightness

$$SB_{\text{Detected}} = \frac{N_{CC}}{P\sqrt{\Delta v_s \Delta v_i}}$$ is about $1.79 \times 10^{-4}$ (s·mW·MHz)$^{-1}$, where $\Delta v_s = 6.4 \times 10^5$ MHZ

and $\Delta v_i = 1.2\times10^5$ MHZ are the filtered bandwidths of the signal photons and the idler photons, respectively. It is worth emphasizing that the comprehensive detection efficiency for the MIR photons in the experiment is 8%, which is higher than the 6.5% quantum efficiency reported in (*41*) and higher than the 2% detection efficiency of the superconducting nanowire single-photon detector (SNSPD) described in the (*33*).

## S4. Quantum Key Distribution (QKD)

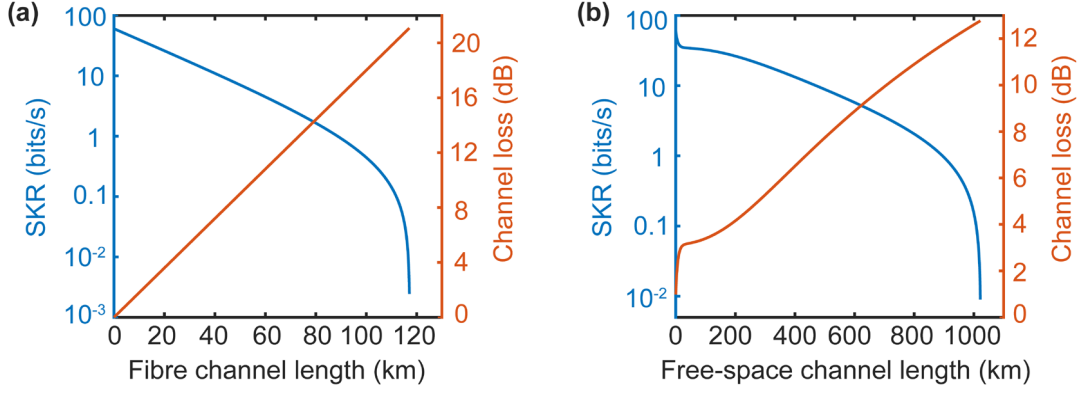

**Fig. S4 Transmission characteristics of QKD systems in optical fiber and free space channels.** (a) The secret key rate (SKR) and channel loss of the QKD system vary with the length of the fiber channel. (b) The secret key rate (SKR) and channel loss of the QKD system vary with the length of the free-space channel.

In order to evaluate the performance of the QKD system in practical applications, we theoretically simulated the transmission characteristics of SKR and quantum bit error rate (QBER) in fiber and free space channels, respectively. The SKR is given by (*17*)

$$R \geq q\{Q[1-f(E)H_2(E)-H_2(E)]\}, \qquad (1.3)$$

where the meanings of the parameters contained in the equation above are the following: $q$ is the base-sift factor ($q = 1/2$ for BBM92); $Q$ is the overall gain, given by

$$Q = 1 - \frac{1-Y_{0A}}{(1+\eta_A\mu/2)^2} - \frac{1-Y_{0B}}{(1+\eta_B\mu/2)^2} + \frac{(1-Y_{0A})(1-Y_{0B})}{(1+\eta_A\mu/2+\eta_B\mu/2-\eta_A\eta_B\mu/2)^2}, \qquad (1.4)$$

$Y_{0A}$ and $Y_{0B}$ are the background count rates on Alice's and Bob's sides, respectively. $\eta_A$ and $\eta_B$ are the detection efficiencies of Alice and Bob, respectively. μ is the average number of photon pairs generated by one pump pulse. $E$ is the overall QBER given by

$$E = e_0 - \frac{(e_0-e_d)\eta_A\eta_B\mu(1+\mu/2)}{Q(1+\eta_A\mu/2)(1+\eta_B\mu/2)(1+\eta_A\mu/2+\eta_B\mu/2-\eta_A\eta_B\mu/2)}, \qquad (1.5)$$

$f(x)$ is the bidirection error correction efficiency. Given the values of our QBER, we take a suitable $f = 1.16$ value (*55*), $H_2(x)$ is the binary entropy function,

$$H_2(x) = -x\log_2(x)-(1-x)\log_2(1-x), \qquad (1.6)$$

Next, we conducted simulations to evaluate the variations in SKR and QBER as functions of the fiber and free space channel lengths based on the experimental parameters of the QKD system presented in Table 1.

**Table 1. Experimental parameters of the QKD system.**

| Repetition rate | $\lambda_s$ | $\lambda_i$ | $\mu$ | $\eta_{0A}$ | $\eta_{0B}$ | $Y_{0A}$ | $e_d$ | $f(x)$ |
|---|---|---|---|---|---|---|---|---|
| 100 MHz | 1555 nm | 3370 nm | 0.03 | 0.042 | 0.0023 | $3\times10^{-5}$ | 1.5% | 1.16 |

$\lambda_s$ and $\lambda_i$ are the wavelengths of the signal and the idler photon, respectively. $\eta_{0A}$ ($\eta_{0B}$) is the detection efficiency in Alice's (Bob's) box, including detector efficiency and internal optical losses. The overall transmittance $\eta_A$ ($\eta_B$) is the product of Alice's (Bob's) channel transmissivity and $\eta_{0A}$ ($\eta_{0B}$). $e_d$ is the intrinsic detector error rate.

**Fiber (Alice's) channel transmissivity** Here, the absorption loss $\alpha$ of the single-mode fiber is set to 0.18 dB/km. Therefore, the transmittance of the optical fiber channel can be expressed as

$$\eta_{fiber} = 10^{-\alpha L/10}, \tag{1.7}$$

where L is the length of the fiber channel. As shown in Figure S4(a), when the free-space channel is at minimum channel loss, the QKD system can still maintain a positive SKR after passing through more than 100 km of communication fiber.

**Free-space (Bob's) channel transmissivity** The calculation of transmissivity $\eta_{space}$ in the free-space channel is relatively complex due to multiple influencing factors, including diffraction loss, the absorption and scattering loss of the atmosphere, and additional losses caused by adverse weather conditions. To simplify the calculation, we ignore the influence of the turbulence and pointing errors. The $\eta_{space}$ can be expressed as (*51*)

$$\eta_{space} = \eta_d \eta_{atm}, \tag{1.8}$$

where $\eta_d$ is the transmissivity due to diffraction given by

$$\eta_d = 1 - e^{-2a_R^2/w_z^2}, \tag{1.9}$$

where $a_R$ (default $a_R$ = 50 cm) is the radius of the receiver telescope, and $w_z$ is the spot size of the beam at the receiver and can be expressed as

$$w_z = w_0 \left(1 + \left(\frac{z\lambda}{\pi w_0^2}\right)^2\right)^{\frac{1}{2}}, \tag{1.10}$$

where $w_0$ (default $w_0$ = 50 cm) is the beam spot size, and $\lambda$ is the wavelength. In our calculations, we neglect the refraction effect of the atmosphere on the elongation of the beam path for different angles. The atmospheric extinction is caused by both aerosol absorption and Rayleigh/Mie scattering, which results in free-space propagation loss. For a fixed altitude $h$ above the ground/sea level, the overall atmospheric transmissivity $\eta_{atm}$ is given by

$$\eta_{atm}(h,z) = e^{-\int_0^z \alpha(h)dz'}, \tag{1.11}$$

where $z$ is the path length in the atmosphere and $h$ is the altitude; for a vertical path, $z$ is assumed to equal $h$. The extinction factor $\alpha(h)$ is given by

$$\alpha(h) = \alpha_0 e^{-h/\bar{h}}, \tag{1.12}$$

where $\alpha_0 \approx 8.1 \times 10^{-5}$ m$^{-1}$ is the extinction factor of 3370 nm idler photons at sea level associated with molecular and aerosol absorption and scattering (*56*). Here, we mainly consider the absorption of mid-infrared photons by water, carbon dioxide, and ozone molecules in the atmosphere; and assume that the scattering is Rayleigh scattering.

**Bob's background noise** The background noise of the free-space channel can be expressed as follows (*51*)

$$Y_{0B} := \eta_{eff}\tilde{n}_B + \tilde{n}_{ex}, \tag{1.13}$$

where $\tilde{n}_B$ is the number of background thermal photons per pulse, $\eta_{eff}$ is the quantum efficiency of the detector, and $\tilde{n}_{ex}$ is extra setup noise (default $\tilde{n}_{ex} = 2 \times 10^{-5}$ photons per pulse). For a receiver with aperture $a_R$, angular field of view $\Omega_{fov}$, and using a detector with time window $\Delta t$ and spectral filter $\Delta\lambda$ around $\lambda$, $\tilde{n}_B$ is given by

$$\tilde{n}_B = \kappa H_\lambda^{sun}\Gamma_R, \quad \Gamma_R := \Delta\lambda\Delta t\Omega_{fov}a_R^2, \tag{1.14}$$

where $H_\lambda^{sun}$ is the solar spectral irradiance. At $\lambda$ = 3370 nm, $H_\lambda^{sun} = 2.85 \times 10^{17}$ photons·m$^{-2}$·s$^{-1}$·nm$^{-1}$·sr$^{-1}$ (*31*). According to the acceptance bandwidth of our up-conversion detection device, we set $\Delta\lambda$ = 5 nm and $\Delta t$ = 1 ns. Assuming $\Omega_{fov}$ = 10$^{-10}$ sr and $a_R$ = 50 cm for the receiving telescope. The dimensionless parameter $\kappa$ has a value of 0.3 for daytime. As shown in Figure S4(b), the QKD system can achieve a transmission distance of more than 1000 km in the free-space channel when the loss of the fiber channel is minimal. The SKR decays rapidly within the 0~30 km transmission range due to the high concentration of water vapor, carbon dioxide, ozone, and tiny particles in this region.

    The communication distance can be further improved by increasing the pump power and optimizing the effective detection efficiency. The above simulation results verify the great potential of this QKD system for long-distance communication in fiber and free-space channels.